\documentclass[12pt]{article}
\usepackage{dsfont,amsthm,natbib,textcomp,amsfonts}
\usepackage[utf8]{inputenc}
\usepackage[colorlinks,citecolor=blue,linkcolor=black,urlcolor=blue]{hyperref}
\usepackage{color}

\usepackage[top=3cm, left=3cm, bottom=3cm, right=3cm]{geometry}
\usepackage{setspace}
\usepackage{graphicx}
\usepackage{bm}
\usepackage{multirow}
\usepackage{dcolumn}
\usepackage{amssymb}
\usepackage{amsbsy}
\usepackage[toc,page]{appendix}

\usepackage[section]{placeins}

\usepackage{sectsty}
\usepackage[normalem]{ulem}

\usepackage[multiple]{footmisc}
\usepackage{caption}

\usepackage{url}
\urldef\finraurl\url{https://www.finra.org/sites/default/files/2019%20Industry%20Snapshot.pdf}
\urldef\bankrupt\url{https://abi-org.s3.amazonaws.com/Newsroom/Bankruptcy_Statistics/Quarterlynonbusinessfilingsbychapter1994-Present.pdf}

\usepackage[figuresleft]{rotating}

\usepackage{booktabs}
\usepackage{threeparttablex}

\usepackage{siunitx}
\sisetup{
detect-mode,
group-digits            = false,
input-symbols           = ( ) [ ] - +,
table-align-text-post   = false,
input-signs             = ,
}   

\def\yyy{%
\bgroup\uccode`\~\expandafter`\string-%
\uppercase{\egroup\edef~{\noexpand\text{\llap{\textendash}\relax}}}%
\mathcode\expandafter`\string-"8000 }

\def\xxxl#1{%
\bgroup\uccode`\~\expandafter`\string#1%
\uppercase{\egroup\edef~{\noexpand\text{\noexpand\llap{\string#1}}}}%
\mathcode\expandafter`\string#1"8000 }

\def\xxxr#1{%
\bgroup\uccode`\~\expandafter`\string#1%
\uppercase{\egroup\edef~{\noexpand\text{\noexpand\rlap{\string#1}}}}%
\mathcode\expandafter`\string#1"8000 }

\def\textsymbols{\xxxl[\xxxr]\xxxl(\xxxr)\yyy}

\makeatletter
\let\estinput\@@input
\makeatother

\newcommand{\estauto}[3]{
\vspace{.75ex}{
\textsymbols
\begin{tabular}{l*{#2}{#3}}
\toprule
\estinput{#1}
\bottomrule
\addlinespace[.75ex]
\end{tabular}
}
}

\newcommand{\Figtext}[1]{%
\begin{tablenotes}[para,flushleft]
#1
\end{tablenotes}
}
\newcommand{\Fignote}[1]{\Figtext{\emph{Note:}~#1}}

\pagenumbering{gobble}

\usepackage{authblk}

\title{%
Financial Adviser Misconduct and Labor Market Penalties: 
Uncovering Racial Disparities in the Absence of Gender Gaps%
\footnote{%
I thank Lothar Gampel for legal assistance regarding the data usage 
and Anna Ulrichshofer for research assistance. 
I also thank Mark Egan for helful comments and sharing information on data construction. 
This work is supported by JSPS KAKENHI Grant Number JP24K04832.
This paper was previously circulated under the title 
``Gender Gaps and Racial Disparities in the Labor Market Penalties for Financial Advisers Misconduct" (2020). 
}
}

\author{%
Jun Honda%
\thanks{%
Faculty of Economics and Law,
Shinshu University, Japan. %
junhonda@shinshu-u.ac.jp. %
}
}

\date{%
\today%
}

\begin{document}

\maketitle

\thispagestyle{empty}

\begin{abstract}

Using a comprehensive matched employer–employee dataset for U.S. financial advisers from 2008 to 2018, we revisit established evidence on labor market penalties following financial misconduct. 
Prior studies report that female advisers are 20\% more likely to exit their firms following misconduct and that similar disparities exist for non-white advisers. 
However, by disaggregating misconduct into distinct disclosure events
-- differentiating those that nearly always trigger job terminations from those that do not -- 
we show that the apparent gender gap vanishes, while significant racial disparities persist. 
Specifically, non-white advisers face approximately 24\% higher job separation rates than their white counterparts. 
Robustness checks confirm these findings across alternative specifications, suggesting that race-based differential treatment in the labor market is a distinct phenomenon warranting further investigation.

\bigskip



\end{abstract}

\pagestyle{empty} 

\tableofcontents 

\cleardoublepage 

\cleardoublepage 
\pagestyle{plain} 
\setcounter{page}{1} 

\break



\section{Introduction}
\label{section: intro}

In the U.S. financial advisory industry
-- where more than 70\% of advisers are male and nearly 90\% are white --
labor market outcomes remain a critical concern, particularly in terms of employment penalties following financial misconduct. 
Prior research \citep*[e.g.,][]{egan2022harry} documents that female advisers face a roughly 20\% higher likelihood of job separation after misconduct, 
and similar disparities exist for non‐white advisers. 
However, these studies typically rely on an aggregated measure of “Misconduct” that combines a variety of disclosure events, 
potentially masking important differences in how various types of misconduct affect employment outcomes.

To address this limitation, our study revisits and replicates the established findings using a comprehensive matched employer–employee dataset constructed from FINRA’s BrokerCheck database covering the period 2008–2018. This unique dataset not only captures detailed compliance records
-- from customer complaints and disputes to regulatory actions, criminal records, and employer terminations --
but also enables us to track the full employment history of every financial adviser from their entry into the industry.

Our approach refines the measurement of financial misconduct by disaggregating disclosure events into distinct categories. We distinguish between events that almost invariably lead to job termination
-- such as employer disciplinary actions --
and those that are less consequential, such as customer disputes. 
This disaggregation is crucial because it allows us to isolate the specific components driving the observed labor market penalties. Our analysis reveals that while the aggregated measure of misconduct replicates the previously documented gender gap, the gap essentially vanishes once we account for the heterogeneous consequences of different disclosure events. 
In stark contrast, racial disparities persist: 
non-white advisers are approximately 24\% more likely to exit their firms and 20\% more likely to receive employer-induced terminations than their white counterparts.

By controlling for firm-branch-year fixed effects and other observable characteristics
-- including tenure, industry experience, and licensing --
the study provides a more nuanced understanding of labor market penalties in this industry. 
In doing so, we offer new evidence on race-based employment discrimination that challenges the conventional narrative of a gender-dominated penalty gap. 
Our findings underscore the importance of refined measurement in uncovering the true extent of differential treatment and have important implications for both policy and future research on employment discrimination.


\subsection{Related Literature}

\paragraph{Gender Gap.} 
Over recent decades, gender disparities have narrowed in areas such as education, labor market participation, and wages 
\citep{blau2000gender, blau2017gender}. 
Yet persistent gaps remain, notably due to career interruptions that impose lasting wage penalties 
\citep{bertrand2010dynamics, goldin2014grand}. 
In the financial advisory industry, 
\citet{egan2022harry}
document a ``gender punishment gap," showing that female advisers face harsher employment penalties following misconduct. 
Their results suggest that this gap is not driven by productivity differences
-- as measured by assets under management or quality ratings --
but may partly stem from in-group favoritism. 
In contrast, our study controls for the composition of disclosure events and uses individual education measures rather than productivity proxies. 
We find that when these factors are accounted for, the gender gap in labor market penalties essentially disappears.

\paragraph{%
Employment
Discrimination.}
Employment discrimination has been widely studied through methods such as audit studies, pseudo-experiments, and correspondence tests
-- mostly focusing on hiring practices 
(see, e.g., 
\citet[][]{darity1998evidence, altonji1999race, lang2012racial, bertrand2017field, neumark2018experimental, lang2020race}%
). 
Although court cases offer direct evidence of discriminatory treatment, their limited scope at the individual level restricts broader analysis. 
Our approach differs by leveraging disclosed terminations in a matched employer-employee dataset, which allows us to examine race-based discrimination in the U.S. financial services industry. 
Despite constraints from missing detailed demographic and performance data, our findings consistently reveal that minority advisers face significantly harsher termination penalties than their white counterparts 
-- even after accounting for firm-branch-year factors and other observables. 
These results challenge the notion that differential treatment is solely driven by statistical discrimination.

\paragraph{Financial Misconduct.}

A growing literature has exploited FINRA’s BrokerCheck database to investigate financial misconduct and its labor market consequences 
\citep[see, e.g.,][]{charoenwongdoes2019,  clifford2021property, cook2020auditors, dimmock2018fraud, dimmock2018real,  egan2019market, GURUN20211218, honigsberg2021deleting, law2019financial}.
Following 
\citet{egan2022harry}, 
our study examines both gender gaps and racial disparities in labor market penalties related to misconduct. 
In parallel with research on the labor market effects of earnings management by top executives 
(\citep[e.g.,][]{agrawal1999management, beneish1999incentives, desai2006reputational, feroz1991financial, karpoff2008consequences}),
our contribution lies in providing additional evidence at the individual level. By refining the measurement of misconduct through the disaggregation of disclosure events, we offer new insights into the distinct impacts of financial misconduct on employment outcomes across gender and race.




\section{Data}
\label{section: data}

We construct a comprehensive, matched employer–employee dataset covering all U.S. financial advisers using FINRA’s BrokerCheck database via the Central Registration Depository (CRD). 
This section details our data on adviser-level employment records, gender and race, the measurement of financial misconduct, and summary statistics.

\subsection{Adviser-Level Data} 
\label{subsec: data FINRA SEC}

Following the methodology of 
\citet{egan2022harry}, 
we create an adviser-year panel dataset from Form U4. 
Form U4 provides detailed information on advisers’ employment histories, regulatory licenses (e.g., industry exams), and compliance records
-- including various disclosure events 
(see 
Appendix \ref{appendix: def disclosure} 
for definitions).

\subsubsection{Employment Record}
\label{subsection: data employment}

The BrokerCheck database offers two employment history records:
\begin{itemize}
\item 
{\bf Registration History:} 
Lists registered securities firms along with unique CRD numbers, firm names, branch addresses, and monthly employment periods.
\item 
{\bf Employment History:} 
Documents an adviser’s work record over the past 10 years, covering both industry and non-industry activities 
(e.g., full-time/part-time work, self-employment, military service, unemployment, and education).
\end{itemize}
For our primary analysis, we rely on Registration History, as done by 
\citet{egan2022harry}.




\subsubsection{Gender and Race}
\label{subsection: gender race}

Since FINRA does not report gender or race, we supplement this information using complementary methods:
\begin{itemize}
\item 
{\bf Gender Identification:} 
We employ the R package gender 
\citep{mullen2018}
to match advisers’ first names with historical Social Security data. 
By imposing an 80\% accuracy threshold, we assign gender to approximately 95\% of advisers, with females comprising around 30\% of the sample.
\item 
{\bf Race Identification:} 
We use the Python package ethnicolr 
\citep{laohaprapanon2019}, 
which classifies advisers into Asian, Black, Hispanic, and white using Florida voting registration data. With a target accuracy of 50\%, this method identifies race for about 99\% of advisers, with non-white advisers representing roughly 12\%. 
Alternative sources yield similar results; 
hence, we collapse race into a binary classification: 
white majority versus non-white minority.
\end{itemize}

%
%
%
%
%
%
%

\subsubsection{Limitations}
\label{subsection: data limitations}

The FINRA BrokerCheck database is subject to survivorship bias, as data completeness depends on an adviser's registration status. 
Brokers active within the past 10 years have more complete records than those deregistered earlier. 
To mitigate this bias, we limit our sample to adviser-year observations from 2008 to 2018, yielding approximately 7.9 million observations covering 1.2 million advisers
-- nearly half of whom have exited the industry.


\subsection{Measurement of Financial Misconduct}
\label{subsec: definition of misconduct}


According to Form U4, there are six broad categories of disclosure events 
(with 23 sub-categories). 
\citet{egan2019market, egan2022harry}
define “Misconduct” by aggregating six types of disclosure events, including: 
(I) Customer Dispute - Settled; 
(II) Employment Separation After Allegations; 
(III) Regulatory - Final; 
(IV) Criminal - Final Disposition; 
(V) Customer Dispute - Award/Judgment; 
and 
(VI) Civil - Final. 
Detailed definitions are provided in 
Appendix \ref{appendix: def disclosure}.

\subsection{Summary Statistics} 
\label{subsec: summary stats}
Table \ref{table: ss 1}
presents summary statistics for financial advisers across gender and race, covering both observable characteristics 
(e.g., industry experience, tenure, job transitions, and regulatory licenses) 
and compliance records (i.e., the annual incidence of disclosure events). 
\begin{table}[t!]
\centering
\setlength{\tabcolsep}{5pt}
\scriptsize 
\caption{%
Summary Statistics by Gender and Race
}
\label{table: ss 1}
\begin{threeparttable}
\estauto{tables/table_1_.tex}{20}{c}
\Fignote{%
Observations are based on adviser–year panel data spanning 2008–2018. 
Advisers are categorized by gender and race (white majority versus non‐white minority; see Section \ref{subsection: gender race} for details). 
The values in ``Licenses/Industry Exams" 
are the percentages of its license holders over all observations.
See the Appendix \ref{appendix: def exam} for the definitions of licenses/qualifications in 
Table \ref{table: ss 1}.A.
Each value in 
Table \ref{table: ss 1}.B
indicates
the annual incidence of the disclosure event 
in 
percentage points, 
which is given by a dummy variable for 
whether an adviser has encountered a disclosure event in the respective category
at least once within a given year. 
}
\end{threeparttable}
\end{table}
Our sample exhibits clear differences: 
male advisers generally have longer experience and tenure and hold more licenses compared to female advisers, and within each gender, observable differences exist between white majority and non-white minority advisers.

\subsection{Data Validity and Comparison} 
\label{subsec: validity}

To ensure the robustness of our analysis, we compare key summary statistics from our dataset with those reported by 
\citet{egan2022harry}. 
Our sample, which spans 2008--2018, exhibits characteristics—such as industry experience, tenure, regulatory license holdings, and the incidence of disclosure events—that are broadly consistent with the findings in 
\citet{egan2022harry}, 
whose data cover 2005–2015. Although minor differences arise
-- likely reflecting post-financial crisis dynamics and sample period variations --
the overall consistency in employment records and compliance measures validates our data construction. 
This comparison reinforces the external validity of our dataset and establishes a reliable foundation for our replication and subsequent analysis of gender and racial disparities in labor market penalties.


\section{%
Replication of 
Prior Findings%
}
\label{section: replications}


In this section, we replicate the key findings from 
\citet{egan2022harry}
regarding labor market penalties following financial misconduct, focusing on both gender and racial differences.

\subsection{%
Replicating Gender Differences in Employment Separations%
}
\label{subsection: gender gap revisit}



We begin by replicating the estimation strategy used by 
\citet{egan2022harry}
to examine the gender penalty in job separations following misconduct. 
The baseline specification is given by:
\begin{eqnarray}
\label{eq: job separation g 1}
\text{Separation}_{iqjlt+1} & = &   \beta_1 \ \text{Female}_{i} 
\ + \ \beta_2 \ \text{Misc}_{iqjlt}
\ + \ \beta_3 \ \text{Misc}_{iqjlt} \ \times \ \text{Female}_{i}  \\
& & \ + \ \beta_4 \ X_{it}
  + \ \mu_{qjlt} \  + \ \varepsilon_{iqjlt}, 
\nonumber
\end{eqnarray}
where 
$\text{Separation}_{iqjlt+1}$ is a dummy variable equal to one if adviser $i$ leaves firm $j$ in year $t+1$.
The variable $\text{Misc}_{iqjlt}$ indicates whether adviser $i$ encounters any misconduct-related disclosure event in year $t$. 
The interaction term $\text{Misc}_{iqjlt} \times \ \text{Female}_{i}$ captures the gender penalty. 
The vector $X_{it}$ includes adviser characteristics such as industry experience, while $\mu_{qjlt}$ represents firm-year-county-license fixed effects.


Table \ref{table: js 1} 
reports our replication estimates. 
\begin{table}[t!]
\centering
\setlength{\tabcolsep}{5pt}
\caption{%
Replication of Gender Differences in Job Separations
Following Misconduct Disclosures
}
\label{table: js 1}
\begin{threeparttable}
\estauto{tables/table_js_1_g_1_1.tex}{12}{c}
\Fignote{%
Observations are based on the adviser-year panel data 
for 
2008--2018. 
The dependent variable is a dummy for whether an adviser exits their firm in the subsequent year. 
``Misc" denotes a dummy variable for any misconduct-related disclosure, and adviser controls include industry experience, tenure, and licensing variables. 
Fixed effects are included at the firm × year × county × license level.
The coefficients are in percentage points. 
Standard errors are in brackets and clustered by firms.  
\\
\text{* $p < 0.10$, ** $p < 0.05$, *** $p < 0.01$.}
}
\end{threeparttable}
\end{table}
The positive and statistically significant coefficient on the interaction term confirms that, using the aggregated measure of misconduct, female advisers are about 20\% more likely to leave their firm than male advisers, 
consistent with \citet{egan2022harry}.

\subsection{%
Replicating Racial Differences in Employment Separations%
}
\label{subsection: racial gap revisit}

Next, we assess racial disparities by modifying the baseline equation, replacing the gender dummy with a minority dummy:
\begin{eqnarray}
\label{eq: job separation r 1}
\text{Separation}_{iqjlt+1} & = &   \beta_1 \ \text{Minority}_{i} 
\ + \ \beta_2 \ \text{Misc}_{iqjlt}
\ + \ \beta_3 \ \text{Misc}_{iqjlt} \  \times \ \text{Minority}_{i}  \\
& & \ + \ \beta_4 \ X_{it}
  + \ \mu_{qjlt} \  + \ \varepsilon_{iqjlt}.
\nonumber
\end{eqnarray}
Here, $\text{Minority}$ equals one if adviser $i$ belongs to the non-white minority group. 

Table \ref{table: js 1 r} 
reports our findings. 
\begin{table}[t!]
\centering
\setlength{\tabcolsep}{12pt}
\caption{%
Replication of Racial Differences in Job Separations
Following Misconduct Disclosures
}
\label{table: js 1 r}
\begin{threeparttable}
\estauto{tables/table_js_1_r_1_3.tex}{12}{c}
\Fignote{%
Observations are based on the adviser-year panel data 
for 
2008--2018. 
The dependent variable is a dummy for whether an adviser exits their firm in the subsequent year. 
``Misc" denotes a dummy variable for any misconduct-related disclosure, and adviser controls include industry experience, tenure, and licensing variables. 
Fixed effects are included at the firm × year × county × license level.
The coefficients are in percentage points. 
Standard errors are in brackets and clustered by firms.  
\\
\text{* $p < 0.10$, ** $p < 0.05$, *** $p < 0.01$.}
}
\end{threeparttable}
\end{table}
The interaction term
$\text{Misconduct} \ \times \ \text{Minority}$
is positive and statistically significant, indicating that minority advisers are roughly 30\% more likely to leave their firm following a misconduct disclosure compared to their white counterparts.
\section{Main Findings}


In this section, we document how the measurement of financial misconduct critically affects the estimated labor market penalties for financial advisers. We first demonstrate that aggregating various disclosure events yields a significant gender gap in job separations. We then show that disaggregating misconduct into its constituent components eliminates the gender gap, while significant racial disparities persist.




\subsection{%
Decomposing Financial Misconduct%
}
\label{subsec: measurement problem}

To better understand the sources of observed labor market penalties, we partition “Misconduct” into two distinct categories:
\begin{itemize}
\item 
{\bf Category A:} 
Disclosure events that almost invariably lead to job termination 
(e.g., Employment Separation After Allegations).
\item 
{\bf Category B:}
Other misconduct-related disclosures 
(e.g., customer disputes, regulatory actions) that do not automatically trigger termination.
\end{itemize}
If female advisers experience a higher proportion of Category A events within the aggregated measure, this composition effect could spuriously inflate the apparent gender penalty. Our data reveal that while the aggregated measure replicates a 20\% higher job separation rate for female advisers, disaggregating the events eliminates the gender difference. In other words, conditional on the type of misconduct, females and males exhibit similar separation rates.

\subsection{%
Absence of a Gender Gap%
}
\label{subsec: absence}

To test this, we consider 
equation 
We test this hypothesis using the following specification:
\begin{eqnarray}
\label{eq: job separation g 2}
\text{Separation}_{iqjlt+1} & = &   \beta_1 \ \text{Female}_{i} 
\ + \ \beta_2 \ \text{Disc}_{iqjlt}
\ + \ \beta_3 \ \text{Disc}_{iqjlt} \ \times \ \text{Female}_{i}  \\
& & \ + \ \beta_4 \ X_{it}
  + \ \mu_{qjlt} \  + \ \varepsilon_{iqjlt}, 
\nonumber
\end{eqnarray}
Table \ref{table: js 1 2}
reports our estimates. 
\begin{table}[t!]
\centering
\setlength{\tabcolsep}{8pt}
\footnotesize 
\caption{%
The Effect of Disaggregated Disclosure Events on Gender Differences in Job Separations
}
\label{table: js 1 2}
\begin{threeparttable}
\estauto{tables/table_js_2_g_1_2.tex}{12}{c}
\Fignote{Observations are based on the adviser-year panel data  over the period 2008-2018. 
The dependent variable is a dummy variable equal to one if an adviser worked for a given firm in 
a given year (excluding the last year 2018)
and left the firm by the end of next year.
The dependent variable is a dummy variable equal to one if an adviser worked for a given firm in 
a given year (excluding the last year 2018)
and left the firm by the end of next year.  
The columns
(1)--(3) 
correspond
to the respective sets of disclosure events: 
(1) Misconduct (defined in Section \ref{subsec: definition of misconduct}),
(2) Misconduct excluding Event (II),
(3) 
Event (II) Employment Separation After Allegations.
The variable ``${\text{\footnotesize Disclosure}}$'' is a dummy for whether an adviser has encountered any disclosure event in the corresponding set, at least once in a given year. 
The coefficients are in percentage points. 
Standard errors are in brackets and clustered by firms.  
\\
\text{* $p < 0.10$, ** $p < 0.05$, *** $p < 0.01$.}
}
\end{threeparttable}
\end{table}
When 
``Disclosure" 
is defined to include all misconduct events, the interaction term is positive and significant. 
However, after excluding Category A events (Employment Separation After Allegations) from the measure, the coefficient on
$\text{Disclosure} \ \times \ \text{Female}$
becomes statistically insignificant. 
This finding indicates that the previously observed gender gap is driven primarily by the composition of disclosure events rather than by an intrinsic gender bias in employer punishment.


In summary, while aggregating disclosure events can improve estimation precision
-- given their infrequent and serially correlated nature -- 
it may also obscure important heterogeneity in their labor market consequences. Accounting for the distinct effects of different events is thus crucial in evaluating the true gender gap in penalties for financial misconduct.



\subsection{%
Persistent Racial Disparities%
} 
\label{subsection: racial job separation 2}
In contrast to the gender findings, racial disparities remain robust. 
We re-estimate the model by replacing the gender dummy with a minority indicator:
\begin{eqnarray}
\label{eq: job separation 3}
\text{Separation}_{iqjlt+1} & = &   \beta_1 \ \text{Minority}_{i} 
\ + \ \beta_2 \ \text{Disc}_{iqjlt}
\ + \ \beta_3 \ \text{Disc}_{iqjlt} \times \ \text{Minority}_{i}  \\
& & \ + \ \beta_4 \ X_{it}
  + \ \mu_{qjlt} \  + \ \varepsilon_{iqjlt}, 
\nonumber
\end{eqnarray}
\begin{table}[t!]
\centering
\setlength{\tabcolsep}{8pt}
\footnotesize 
\begin{threeparttable}
\caption{%
Racial Disparities in Job Separations: Impact of Disaggregated Misconduct Measures
}
\label{table: js 2 r}


\estauto{tables/table_js_2_r_1_2.tex}{12}{c}
\Fignote{%
The dependent variable is a dummy variable equal to one if an adviser worked for a given firm in 
a given year (excluding the last year 2018)
and left the firm by the end of next year.  
The columns
(1)--(3) 
correspond
to the respective sets of disclosure events: 
(1) Misconduct (defined in Section \ref{subsec: definition of misconduct}),
(2) Misconduct excluding Event (II),
(3) 
Event (II) Employment Separation After Allegations.
The variable ``${\text{\footnotesize Disclosure}}$'' is a dummy for whether an adviser has encountered any disclosure event in the corresponding set, at least once in a given year. 
The coefficients are in percentage points. 
Standard errors are in brackets and clustered by firms.  
\\
\text{* $p < 0.10$, ** $p < 0.05$, *** $p < 0.01$.}
}
\end{threeparttable}
\end{table}
As shown in 
Table \ref{table: js 2 r}, 
the interaction term
``$\text{Disclosure} \ \times \ \text{Minority}$"
remains positive and statistically significant even when we exclude Category A events from the misconduct measure. Minority advisers are approximately 24\% more likely to exit their firms following a misconduct disclosure, and this effect is robust to alternative specifications. 
Moreover, further analysis indicates that minority advisers are about 20\% more likely to receive employer-induced terminations relative to their white counterparts.


\subsection{%
Implications of the Findings
} 
\label{subsection: implication}

Our findings underscore a critical insight: while the apparent gender penalty in job separations can be attributed to the aggregation of heterogeneous disclosure events, the racial disparities in employment outcomes persist even after accounting for these measurement issues. 
This suggests that race-based differential treatment in the labor market is a distinct phenomenon that warrants further investigation.


\section{%
Robustness Checks and Additional Analyses
}
\label{section: robustness}


To further validate our findings, we perform a series of robustness checks addressing alternative explanations and potential confounding factors.


\subsection{Disclosure Interdependence}
\label{subsection: robustness Interdependence}

We investigate whether gender- or race-specific correlations between different types of disclosure events contribute to the observed punishment gaps. Specifically, we test if the sequencing of events—such as a non-termination disclosure (e.g., customer dispute) triggering a subsequent employer disciplinary action (Event (II))
-- differs by gender or race. Our analysis shows that while there is a moderate overall positive correlation between these events, there is no significant female-specific or minority-specific positive correlation. 
This confirms that the observed gender gap largely stems from the composition of disclosure events rather than from differential sequencing.


\subsection{Alternative Specifications}
\label{subsection: robustness alternative}

We further assess the sensitivity of our results using alternative model specifications and additional controls:
\begin{itemize}
\item 
{\bf Settlement Costs:}
We examine whether the cost associated with disclosure events differs between minority and majority groups. Our findings indicate no significant difference in the associated costs.
\item
{\bf Industry Experience and Tenure:}
Our main results hold across various career stages, and they remain robust when we restrict the sample to “loyal” advisers who have never switched firms before termination.
\item 
{\bf Education:}
To address concerns that lower average educational attainment among minority advisers might bias the estimated termination gap, we restrict the sample to advisers with at least a university-level education. Even with this constraint, minority advisers are still at least 25\% more likely to be terminated than their white counterparts. 
Note that we cannot control for pre-market factors, a limitation compared to studies such as \citet{neal1996role}.
\end{itemize}

Together, these robustness checks reinforce our primary conclusions: after properly disaggregating financial misconduct, the gender gap in labor market penalties disappears, while significant racial disparities persist.



\section{Concluding Remarks}
\label{section: conclusion}

In this paper, we reexamine labor market penalties for financial misconduct in the U.S. financial advisory industry using a comprehensive matched dataset from FINRA’s BrokerCheck database covering 2008–2018. Our replication of prior evidence confirms that, when misconduct is measured in aggregate, female advisers appear to face a higher likelihood of job separation following misconduct. However, by disaggregating misconduct into its constituent disclosure events, we find that the apparent gender gap is driven primarily by compositional effects and essentially disappears. In stark contrast, significant racial disparities persist: non‐white advisers are approximately 24\% more likely to exit their firms following misconduct than their white counterparts. 
Robustness checks
-- examining alternative model specifications and controlling for key firm and individual characteristics --
further validate these findings. 

These results underscore that, within the financial advisory industry, differential labor market outcomes are driven predominantly by race-based factors rather than by gender bias. An important avenue for future research is to investigate the underlying mechanisms that sustain these disparities. In particular, exploring the role of in-group favoritism
-- as suggested by 
\citet{egan2022harry} --
may shed light on how the composition of management and leadership within firms influences the severity of employment penalties across different gender and racial groups. 
Understanding whether employers exhibit leniency toward individuals who share their demographic characteristics could provide valuable insights into the persistence of racial disparities in employment outcomes and inform policy interventions aimed at promoting fairer treatment in the labor market.


\newpage 









\setcounter{figure}{0}
\renewcommand{\thefigure}{\Alph{section}.\arabic{figure}}


\setcounter{table}{0}
\renewcommand{\thetable}{\Alph{section}.\arabic{table}}





%


\appendix

\part*{Appendix}


\section{Definition of the Major Disclosure Events}
\label{appendix: def disclosure}

Disclosure events details are described in Form U4.%
\footnote{%
The Form U4 is available via \url{https://www.finra.org/sites/default/files/form-u4.pdf} 
(accessed February 20, 2025).
Note that the definition of each event is given in the FINRA's BrokerCheck report for financial advisers (registered representatives) who have indeed received that disclosure in the past. 
See 
\url{https://brokercheck.finra.org/} 
and also \url{https://www.finra.org/sites/default/files/AppSupportDoc/p015111.pdf}
(accessed February 20, 2025).
} 
Below we consider the major disclosure events 
excluding 
those on appeal and pending ones,
and give their definitions used in the FINRA's BrokerCheck database.%


{\footnotesize

\vspace{0.5cm} \noindent {\bf Customer Dispute - Settled:}
This type of disclosure event involves a consumer-initiated, investment-related complaint, arbitration proceeding or civil
suit containing allegations of sale practice violations against the broker that resulted in a monetary settlement to the
customer.

\vspace{0.5cm} \noindent {\bf Customer Dispute - Award / Judgment:}
This type of disclosure event involves a final, consumer-initiated, investment-related arbitration or civil suit containing
allegations of sales practice violations against the broker that resulted in an arbitration award or civil judgment for the
customer.

\vspace{0.5cm} \noindent {\bf Customer Dispute - Closed-No Action / Withdrawn / Dismissed / Denied:}
This type of disclosure event involves (1) a consumer-initiated, investment-related arbitration or civil suit containing
allegations of sales practice violations against the individual broker that was dismissed, withdrawn, or denied; or (2) a
consumer-initiated, investment-related written complaint containing allegations that the broker engaged in sales practice
violations resulting in compensatory damages of at least \$5,000, forgery, theft, or misappropriation, or conversion of funds
or securities, which was closed without action, withdrawn, or denied.

%
%
%
%

\vspace{0.5cm} \noindent {\bf Criminal - Final Disposition:}
This type of disclosure event involves a conviction or guilty plea for any felony or certain misdemeanor offenses, including
bribery, perjury, forgery, counterfeiting, extortion, fraud, and wrongful taking of property that is currently on appeal.

\noindent
{\bf Type:} Felony, Misdemeanor.

%
%
%
%

\vspace{0.5cm} \noindent {\bf Civil - Final:}
This type of disclosure event involves (1) an injunction issued by a court in connection with investment-related activity, (2)
a finding by a court of a violation of any investment-related statute or regulation, or (3) an action brought by a state or
foreign financial regulatory authority that is dismissed by a court pursuant to a settlement agreement.

%
%
%
%
%

\vspace{0.5cm} \noindent {\bf Employment Separation After Allegations:}
This type of disclosure event involves a situation where the broker voluntarily resigned, was discharged, or was permitted
to resign after being accused of (1) violating investment-related statutes, regulations, rules or industry standards of
conduct; (2) fraud or the wrongful taking of property; or (3) failure to supervise in connection with investment-related
statutes, regulations, rules, or industry standards of conduct.

\noindent
{\bf Termination Type:} Discharged, Permitted to Resign, Voluntary Resignation.

\vspace{0.5cm} \noindent {\bf Regulatory Final:}
This type of disclosure event may involve (1) a final, formal proceeding initiated by a regulatory authority (e.g., a state
securities agency, self-regulatory organization, federal regulatory such as the Securities and Exchange Commission,
foreign financial regulatory body) for a violation of investment-related rules or regulations; or (2) a revocation or
suspension of a broker's authority to act as an attorney, accountant, or federal contractor.

%
%
%
%
%
%
%

\vspace{0.5cm} \noindent {\bf Financial - Final:}
This type of disclosure event involves a bankruptcy, compromise with one or more creditors, or Securities Investor
Protection Corporation liquidation involving the broker or an organization/brokerage firm the broker controlled that
occurred within the last 10 years.

\noindent
{\bf Action Type:} 
Bankruptcy [Chapter 7, Chapter 11, Chapter 13, Other],
Compromise, Declaration, Liquidation, Receivership, Other.

\noindent
{\bf Disposition Type:} 
Direct Payment Procedure, Discharged, Dismissed, Dissolved, SIPA Trustee Appointed, Satisfied/Released, Other.

%
%
%
%
%

\vspace{0.5cm} \noindent {\bf Judgment / Lien:}
This type of disclosure event involves an unsatisfied and outstanding judgments or liens against the broker.

\noindent
{\bf Type:} Civil, Tax.

\vspace{0.5cm} \noindent {\bf Civil Bond:}
This type of disclosure event involves a civil bond for the broker that has been denied, paid, or revoked by a bonding
company.

\vspace{0.5cm} \noindent {\bf Investigation:}
This type of disclosure event involves any ongoing formal investigation by an entity such as a grand jury state or federal
agency, self-regulatory organization or foreign regulatory authority. Subpoenas, preliminary or routine regulatory inquiries,
and general requests by a regulatory entity for information are not considered investigations and therefore are not
included in a BrokerCheck report.

}


\newpage


\section{Definition of the Major Qualification Exams (Licenses)}
\label{appendix: def exam}

The definitions of qualification exams (licenses) are described in the FINRA website.%
\footnote{See the website: 
\url{https://www.finra.org/registration-exams-ce/qualification-exams}
(accessed February 20, 2025).
} 
Below we consider the major qualification exams (Series 6, 7, 24, 63, 65, 66) as in the main text and give their definitions used in the website.
Series 6 and 7 are categorized as ``FINRA Representative-level Exams", 
Series 24 as ``FINRA Principal-level Exams",  
Series 63, 65, and 66 as ``North American Securities Administrators Association (NASAA) Exams". 
Note that the definitions of NASAA Exams are given by the NASAA website.%
\footnote{See the website: 
\url{https://www.nasaa.org/exams/exam-study-guides/}
(accessed February 20, 2025).
}

{\footnotesize

\vspace{0.5cm} \noindent {\bf Series 6:}
The Series 6 exam -- the Investment Company and Variable Contracts Products Representative Qualification Examination (IR) -- assesses the competency of an entry-level representative to perform their job as an investment company and variable contracts products representative.
The exam measures the degree to which each candidate possesses the knowledge needed to perform the critical functions of an investment company and variable contract products representative, including sales of mutual funds and variable annuities.

\vspace{0.5cm} \noindent {\bf Series 7:}
The Series 7 exam -- the General Securities Representative Qualification Examination (GS) -- assesses the competency of an entry-level registered representative to perform their job as a general securities representative. 
The exam measures the degree to which each candidate possesses the knowledge needed to perform the critical functions of a general securities representative, including sales of corporate securities, municipal securities, investment company securities, variable annuities, direct participation programs, options and government securities.

\vspace{0.5cm} \noindent {\bf Series 24:}
The Series 24 exam -- the General Securities Principal Qualification Exam (GP) -- assesses the competency of an entry-level principal to perform their job as a principal dependent on their corequisite registrations.
The exam measures the degree to which each candidate possesses the knowledge needed to perform the critical functions of a principal, including the rules and statutory provisions applicable to the supervisory management of a general securities broker-dealer.%
\footnote{In addition to the Series 24 exam, candidates must pass the Securities Industry Essentials (SIE) Exam (since October 1, 2018 with a complete overhaul) and a representative-level qualification exam, or the Supervisory Analysts Exam (Series 16) exam, to hold an appropriate principal registration. 
See the FINRA website for the definitions of related exams. }

\vspace{0.5cm} \noindent {\bf Series 63:}
The Series 63 exam -- the Uniform Securities State Law Examination -- is a North American Securities Administrators Association (NASAA) exam administered by FINRA.


\noindent	
(Definition given by NASAA:)
The Uniform Securities Agent State Law Examination was developed by NASAA in cooperation with representatives of the securities industry and industry associations. The examination, called the Series 63 exam, is designed to qualify candidates as securities agents. The examination covers the principles of state securities regulation reflected in the Uniform Securities Act (with the amendments adopted by NASAA and rules prohibiting dishonest and unethical business practices). The examination is intended to provide a basis for state securities administrators to determine an applicant?s knowledge and understanding of state law and regulations.

\vspace{0.5cm} \noindent {\bf Series 65:}
The Series 65 exam -- the NASAA Investment Advisers Law Examination -- is a North American Securities Administrators Association (NASAA) exam administered by FINRA.


\noindent	
(Definition given by NASAA:)
The Uniform Investment Adviser Law Examination and the available study outline were developed by NASAA. The examination, called the Series 65 exam, is designed to qualify candidates as investment adviser representatives. The exam covers topics that have been determined to be necessary to understand in order to provide investment advice to clients.

\vspace{0.5cm} \noindent {\bf Series 66:}
The Series 66 exam -- the NASAA Uniform Combined State Law Examination -- is a North American Securities Administrators Association (NASAA) exam administered by FINRA.


\noindent	
(Definition given by NASAA:)
The Uniform Combined State Law Examination was developed by NASAA based on industry requests. The examination (also called the ``Series 66") is designed to qualify candidates as both securities agents and investment adviser representatives. The exam covers topics that have been determined to be necessary to provide investment advice and effect securities transactions for clients.%
\footnote{	
	The FINRA Series 7 is a corequisite exam that needs to be successfully completed in addition to the Series 66 exam before a candidate can apply to register with a state.
}

}


\newpage

\bibliographystyle{ier}
\bibliography{reference/bibtex_feb_2025} 

\begin{thebibliography}{28}
\newcommand{\enquote}[1]{``#1''}
\providecommand{\natexlab}[1]{#1}

\bibitem[{Agrawal et~al.(1999)Agrawal, Jaffe and Karpoff}]{agrawal1999management}
\textsc{Agrawal, A., J.~F. Jaffe and J.~M. Karpoff}, \enquote{Management turnover and governance changes following the revelation of fraud,} \emph{The Journal of Law and Economics} 42 (1999), 309--342.

\bibitem[{Altonji and Blank(1999)}]{altonji1999race}
\textsc{Altonji, J.~G. and R.~M. Blank}, \enquote{Race and gender in the labor market,} \emph{Handbook of Labor Economics} 3 (1999), 3143--3259.

\bibitem[{Beneish(1999)}]{beneish1999incentives}
\textsc{Beneish, M.~D.}, \enquote{Incentives and penalties related to earnings overstatements that violate GAAP,} \emph{The Accounting Review} 74 (1999), 425--457.

\bibitem[{Bertrand and Duflo(2017)}]{bertrand2017field}
\textsc{Bertrand, M. and E.~Duflo}, \enquote{Field Experiments on Discriminationa,} in \emph{Handbook of Economic Field Experiments}volume~1 (Elsevier, 2017), 309--393.

\bibitem[{Bertrand et~al.(2010)Bertrand, Goldin and Katz}]{bertrand2010dynamics}
\textsc{Bertrand, M., C.~Goldin and L.~F. Katz}, \enquote{Dynamics of the gender gap for young professionals in the financial and corporate sectors,} \emph{American Economic Journal: Applied Economics} 2 (2010), 228--55.

\bibitem[{Blau and Kahn(2000)}]{blau2000gender}
\textsc{Blau, F.~D. and L.~M. Kahn}, \enquote{Gender differences in pay,} \emph{Journal of Economic Perspectives} 14 (2000), 75--99.

\bibitem[{Blau and Kahn(2017)}]{blau2017gender}
---\hspace{-.1pt}---\hspace{-.1pt}---, \enquote{The gender wage gap: Extent, trends, and explanations,} \emph{Journal of Economic Literature} 55 (2017), 789--865.

\bibitem[{Charoenwong et~al.(2019)Charoenwong, Kwan and Umar}]{charoenwongdoes2019}
\textsc{Charoenwong, B., A.~Kwan and T.~Umar}, \enquote{Does Regulatory Jurisdiction Affect the Quality of Investment-Adviser Regulation?,} \emph{American Economic Review} 109 (2019), 3681--3712.

\bibitem[{Clifford and Gerken(2021)}]{clifford2021property}
\textsc{Clifford, C.~P. and W.~C. Gerken}, \enquote{Property rights to client relationships and financial advisor incentives,} \emph{The Journal of Finance} 76 (2021), 2409--2445.

\bibitem[{Cook et~al.(2020)Cook, Kowaleski, Minnis, Sutherland and Zehms}]{cook2020auditors}
\textsc{Cook, J., Z.~T. Kowaleski, M.~Minnis, A.~Sutherland and K.~M. Zehms}, \enquote{Auditors are known by the companies they keep,} \emph{Journal of Accounting and Economics} (2020), 101314.

\bibitem[{Darity and Mason(1998)}]{darity1998evidence}
\textsc{Darity, W.~A. and P.~L. Mason}, \enquote{Evidence on discrimination in employment: Codes of color, codes of gender,} \emph{Journal of Economic Perspectives} 12 (1998), 63--90.

\bibitem[{Desai et~al.(2006)Desai, Hogan and Wilkins}]{desai2006reputational}
\textsc{Desai, H., C.~E. Hogan and M.~S. Wilkins}, \enquote{The reputational penalty for aggressive accounting: Earnings restatements and management turnover,} \emph{The Accounting Review} 81 (2006), 83--112.

\bibitem[{Dimmock et~al.(2018{\natexlab{a}})Dimmock, Gerken and Graham}]{dimmock2018fraud}
\textsc{Dimmock, S.~G., W.~C. Gerken and N.~P. Graham}, \enquote{Is Fraud Contagious? Coworker Influence on Misconduct by Financial Advisors,} \emph{The Journal of Finance} 73 (2018{\natexlab{a}}), 1417--1450.

\bibitem[{Dimmock et~al.(2018{\natexlab{b}})Dimmock, Gerken and Van~Alfen}]{dimmock2018real}
\textsc{Dimmock, S.~G., W.~C. Gerken and T.~D. Van~Alfen}, \enquote{Real Estate Shocks and Financial Advisor Misconduct,}  (2018{\natexlab{b}}).

\bibitem[{Egan et~al.(2019)Egan, Matvos and Seru}]{egan2019market}
\textsc{Egan, M., G.~Matvos and A.~Seru}, \enquote{The market for financial adviser misconduct,} \emph{Journal of Political Economy} 127 (2019), 233--295.

\bibitem[{Egan et~al.(2022)Egan, Matvos and Seru}]{egan2022harry}
---\hspace{-.1pt}---\hspace{-.1pt}---, \enquote{When Harry fired Sally: The double standard in punishing misconduct,} \emph{Journal of Political Economy} 130 (2022), 1184--1248.

\bibitem[{Feroz et~al.(1991)Feroz, Park and Pastena}]{feroz1991financial}
\textsc{Feroz, E.~H., K.~Park and V.~S. Pastena}, \enquote{The financial and market effects of the SEC's accounting and auditing enforcement releases,} \emph{Journal of Accounting Research} 29 (1991), 107--142.

\bibitem[{Goldin(2014)}]{goldin2014grand}
\textsc{Goldin, C.}, \enquote{A grand gender convergence: Its last chapter,} \emph{American Economic Review} 104 (2014), 1091--1119.

\bibitem[{Gurun et~al.(2021)Gurun, Stoffman and Yonker}]{GURUN20211218}
\textsc{Gurun, U.~G., N.~Stoffman and S.~E. Yonker}, \enquote{Unlocking clients: The importance of relationships in the financial advisory industry,} \emph{Journal of Financial Economics} 141 (2021), 1218--1243.

\bibitem[{Honigsberg and Jacob(2021)}]{honigsberg2021deleting}
\textsc{Honigsberg, C. and M.~Jacob}, \enquote{Deleting misconduct: The expungement of BrokerCheck records,} \emph{Journal of Financial Economics} 139 (2021), 800--831.

\bibitem[{Karpoff et~al.(2008)Karpoff, Lee and Martin}]{karpoff2008consequences}
\textsc{Karpoff, J.~M., D.~S. Lee and G.~S. Martin}, \enquote{The consequences to managers for financial misrepresentation,} \emph{Journal of Financial Economics} 88 (2008), 193--215.

\bibitem[{Lang and Lehmann(2012)}]{lang2012racial}
\textsc{Lang, K. and J.-Y.~K. Lehmann}, \enquote{Racial discrimination in the labor market: Theory and empirics,} \emph{Journal of Economic Literature} 50 (2012), 959--1006.

\bibitem[{Lang and Spitzer(2020)}]{lang2020race}
\textsc{Lang, K. and A.~K.-L. Spitzer}, \enquote{Race Discrimination: An Economic Perspective,} \emph{Journal of Economic Perspectives} 34 (2020), 68--89.

\bibitem[{Laohaprapanon and Sood(2019)}]{laohaprapanon2019}
\textsc{Laohaprapanon, S. and G.~Sood}, \emph{ethnicolr: Predict Race and Ethnicity From Name} (2019), python package version 0.3.0.

\bibitem[{Law and Mills(2019)}]{law2019financial}
\textsc{Law, K.~K. and L.~F. Mills}, \enquote{Financial gatekeepers and investor protection: Evidence from criminal background checks,} \emph{Journal of Accounting Research} 57 (2019), 491--543.

\bibitem[{Mullen(2018)}]{mullen2018}
\textsc{Mullen, L.}, \emph{gender: Predict Gender from Names Using Historical Data} (2018), r package version 0.5.2.

\bibitem[{Neal and Johnson(1996)}]{neal1996role}
\textsc{Neal, D.~A. and W.~R. Johnson}, \enquote{The role of premarket factors in black-white wage differences,} \emph{Journal of Political Economy} 104 (1996), 869--895.

\bibitem[{Neumark(2018)}]{neumark2018experimental}
\textsc{Neumark, D.}, \enquote{Experimental research on labor market discrimination,} \emph{Journal of Economic Literature} 56 (2018), 799--866.

\end{thebibliography}

\end{document}